# Anticipatory Digital Twins for Online Head-and-Neck Adaptive Proton Therapy via Foundation-Model Registration

Yizhou Wu[1], Yuheng Li[1], Xiaofeng Yang[2*], Chih-Wei Chang[1*]

[1] Department of Radiation Oncology and Winship Cancer Institute, Emory University, Atlanta, GA 30308, USA
[2] Department of Radiation and Cellular Oncology, University of Chicago, Chicago, IL 60637, USA
xfyang@uchicago.edu (XY); chih-wei.chang@emory.edu (CC)

**Abstract.** Head-and-neck (HN) proton therapy is highly sensitive to anatomical change over a 4-to-6-week course, as tumor shrinkage, weight loss, and setup variation can misposition the Bragg peak near critical organs such as the parotids, oral cavity, brainstem, and spinal cord, leading to target underdosing or organ-at-risk overdosing. Online adaptive proton therapy replans on the anatomy of the day, yet standard workflows rely on offline replanning that requires repeated CT acquisition and roughly a week of preparation, adding burden, cost, and delay. We investigate whether a patient's treatment-day anatomy can be predicted before image acquisition by transferring longitudinal change from a population database. We propose a digital-twin framework built on a pretrained foundation-model deformable registration network used without patient-specific training. A first registration aligns a prior patient's planning CT to the target and carries the prior's during-treatment quality assurance CT (QACT) into the target frame; a second registration estimates the prior's planning-to-QACT change, which is then applied to the target's own planning CT to synthesize predicted CTs (pdCTs) with propagated contours. Using 88 HN patients, each with a planning CT and three QACTs, we show that pdCTs better match treatment-day anatomy than the static planning CT. Compared with the planning CT alone, normalized cross-correlation improves by 22.8%, Dice for organs-at-risk by 20.2%, and CT-number error decreases by 23.4%. Gains are largest for patients with major anatomical change and negligible when anatomy is stable. This cross-patient motion transfer leverages the digital-twin concept to anticipate treatment-day anatomy, enabling personalized online adaptive proton therapy without repeated imaging.

## 1 Introduction

Proton radiotherapy for head-and-neck (HN) cancer is typically delivered in 30-35 fractions over six to seven weeks, during which patient anatomy can change substantially:

primary tumors and nodal masses regress, patients lose weight, and the parotid glands shrink and migrate medially [1]. These changes degrade the planned dose distribution, increasing dose to organs at risk (OARs) and risking target underdosage [2]. Adaptive proton therapy (APT) addresses this by replanning on the treatment-day anatomy [3], but it requires acquiring a new CT scan and re-delineating targets and OARs before a new plan can be generated. The combined burden of image acquisition, contouring, and patient scheduling means APT is usually performed offline, along a five-to-seven-day replanning care path. Such a reactive workflow can compromise treatment quality, because the anatomical drift it responds to has already occurred. Online APT instead aims to capture same-day variation while the patient remains on the couch; however, the complexity of robust proton optimization means more than an hour is typically needed before a new plan is ready for physician and physicist review.

A digital-twin (DT) strategy, which simulates the patient *in silico* to support treatment decisions, offers a path to fast online APT that proactively anticipates treatment-day anatomy and preserves personalized plan quality [4-11]. Following the consensus framing of medical digital twins [12, 13], such a model couples a virtual patient representation to its physical counterpart, synchronizes with treatment-day data, and feeds predictions back to inform adaptation. If a patient's potential treatment-day anatomy can be predicted before imaging, multiple predicted CT (pdCT) image sets spanning plausible anatomical variations can be synthesized offline, and a precomputed DT-based plan library can be constructed to enable fast online APT through optimal plan selection on the treatment-day anatomy. The central difficulty is that anatomical change is patient-specific and partly stochastic, so the generated pdCT sets must be sufficiently diverse to cover this uncertainty.

Existing planning-image tools do not predict future anatomy. Deformable image registration (DIR) propagates contours and accumulates dose, but it operates between already-acquired images and cannot synthesize anatomy that does not yet exist [14, 15]. Synthetic-CT methods (e.g., from CBCT or MRI) reconstruct treatment-day anatomy from a same-day scan, again requiring an acquisition [16, 17]. Statistical and biomechanical models can extrapolate change but typically depend on population training data or patient-specific tumor models [18]. Atlas-based approaches transfer information across patients, but predominantly for segmentation rather than for synthesizing a future CT [19].

In this work, we propose a DT framework that leverages a foundation model to synthesize pdCT with propagated contours for fast HN online APT, via a two-step DIR scheme producing two deformation vector fields ($DVF_1$ and $DVF_2$). The framework makes three contributions. (i) We synthesize multiple pdCT by cross-patient transfer of anatomical change: a prior patient's planning-to-QACT change is aligned to a new target via a cross-patient registration ($DVF_1$) and then re-estimated and applied on the *target*'s own planning CT ($DVF_2$), so the prediction reflects the prior patient's change while remaining anchored to the target's own anatomy. (ii) We integrate a pretrained foundation model into the DT framework, reused without patient-specific training, to provide robust DIR and consistent propagation of target and OAR contours. (iii) We show that a composite score combining intensity, structural, and deformation-regularity metrics can evaluate pdCT quality and feasibility for downstream online HN APT across 1,740 ($DVF_1$) and 5,220 ($DVF_2$) cross-patient pairings, and that improvements over a no-adaptation baseline are largest precisely for the patients whose anatomy changed most.

## 2 Materials and Methods

**Overview.** Fig. 1 summarizes the pipeline. The *target* patient has a planning CT $I_t$ (the TPCT, with planning contours $S_t$); its three during-treatment QA CTs $Q_t^{(1)}$, $Q_t^{(2)}$, $Q_t^{(3)}$ capture anatomy on treatment days and serve as ground truth for evaluation only. A prior-patient library provides previously treated patients, each with a planning CT $I_p$ and three QA CTs $Q_p^{(k)}$, whose planning-to-QACT differences encode that prior patient's anatomical change during treatment. We rank prior patients by image similarity to the target (Sec. 2.1) and then transfer the selected prior patient's planning-to-QACT change onto the target through two deformable registrations (Sec. 2.3), synthesizing predicted CTs $\hat{I}_t^{(k)}$ (pdCTs) with propagated contours $\hat{S}_t^{(k)}$. The pdCT is expected to match the target's treatment-day QACT more closely than the static planning CT does, anticipating anatomical change before any treatment-day image is acquired. Throughout, $R$(F, M) denotes a single pretrained 2D deformable registration model, reused without patient-specific fine-tuning, that returns the DVF $\varphi$ aligning moving image $M$ to fixed image $F$; $\varphi \circ X$ denotes warping image (or contour set) $X$ by $\varphi$.

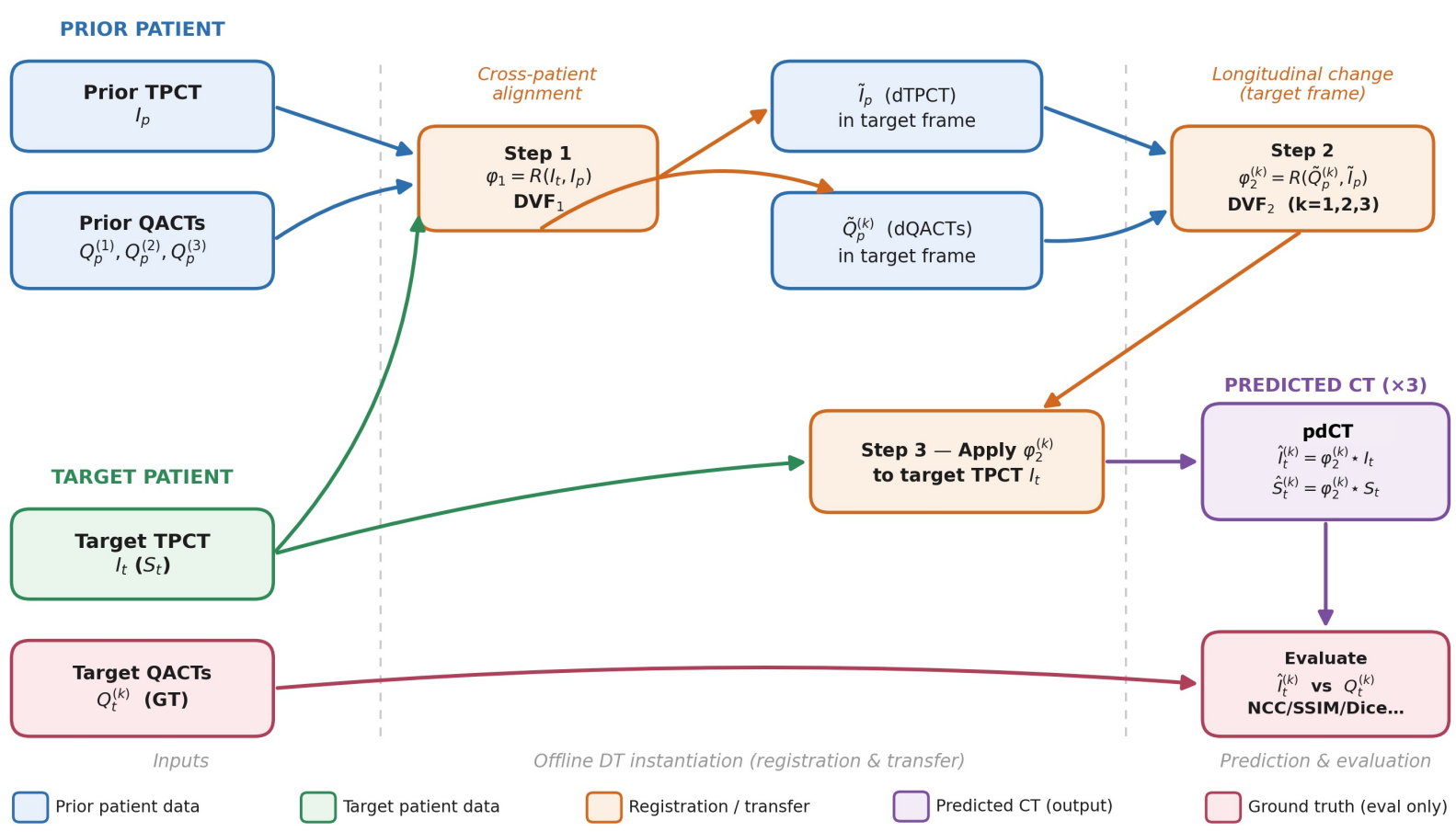

**Fig. 1.** Proposed digital-twin framework for predicted-CT generation. A prior patient's planning CT is aligned to the target ($\varphi_1$, $DVF_1$), bringing the prior QACTs into the target frame; the prior's planning-to-QACT change is estimated per QACT ($\varphi_2^{(k)}$, $DVF_2$) and applied to the target planning CT to synthesize three pdCTs, with contours propagated by the same fields.

### 2.1 Prior patient selection and atlas ranking

All scans are rigidly aligned to the target's planning grid. For each candidate prior patient we compute a composite image-similarity score between $I_t$ and $I_p$ that combines, with equal weight, normalized cross-correlation (NCC) for intensity agreement, learned perceptual image-patch similarity (LPIPS) for deep-feature structural similarity, and mutual information (MI) for statistical dependence. Equal weighting avoids over-favoring any single criterion absent a priori evidence of metric dominance. Candidate priors are ranked by this composite score and the top-$K$ are retained as motion sources; the most similar prior provides the primary prediction.

### 2.2 Pretrained registration foundation model

All registrations $R(\cdot,\cdot)$ are computed with a single, frozen, patient-agnostic foundation model. We use FlexiCT-2D [20], a Vision Transformer pretrained on 2D CT slices by agglomerative continual self-supervised pretraining (patch size 8, embedding dimension 864, 16 blocks, 12 heads); the public checkpoint is reused without fine-tuning. Each volume is encoded slice-wise on an 80×70 patch-token grid; tokens from the last four layers form a 3456-dimensional per-token descriptor, projected to 24 channels by low-rank PCA. Registration is two-stage: a coupled-convex search at 6× downsampling gives a smooth coarse field (±4-grid-point window, inverse-consistency refinement), from which a dense voxel-displacement field $\varphi \in \mathbb{R}^{H\times W\times D\times 3}$ is refined for 500 Adam iterations (lr = 6) on the features (downsampled 2×), with $H$, $W$, $D$ the image height, width, and depth and the last dimension the three displacement components. The objective is a sum-of-squared-differences feature loss plus a first-order diffusion penalty weighted by $\lambda$ ($\lambda = 1$). The output of $\lambda$ ($\lambda = 1$). The output of $R$ is $\varphi$ in voxel units; image and label warps use trilinear and nearest-neighbour interpolation. FlexiCT-2D was pretrained on 266,227 public CT volumes from 56 datasets spanning multiple body regions including HN CT sets [21, 22], and is applied frozen, with no fine-tuning on our cohort.

### 2.3 Cross-patient longitudinal motion transfer

The transfer proceeds in two registration steps followed by prediction.
**Step 1: Cross-patient alignment ($\varphi_1$, DVF$_1$).** A cross-patient registration aligns the prior planning CT to the target planning CT and brings all prior time points onto the target grid:

$$\varphi_1 = R(I_t, I_p),\quad \tilde{I}_p = \varphi_1 \circ I_p,\quad \tilde{Q}_p^{(k)} = \varphi_1 \circ Q_p^{(k)},\quad \text{k} = 1,2,3 \tag{1}$$

where $\tilde{I}_p$ (dTPCT) and $\tilde{Q}_p^{(k)}$ (dQACTs) are the prior patient's planning and QA images expressed in the target's anatomy.
**Step 2: Longitudinal change ($\varphi_2^{(k)}$, DVF$_2$).** On the target grid, a longitudinal registration estimates the prior patient's planning-to-QACT deformation for each of its three QACTs, now expressed in the target's anatomy:

$$\varphi_2^{(\text{k})} = R(\tilde{Q}_p^{(k)}, \tilde{I}_p),\quad \text{k} = 1,2,3 \tag{2}$$

**Step 3: Prediction.** Each longitudinal change is applied to the target patient's own planning CT to synthesize three pdCTs, with the planning contours propagated by the same fields (nearest-neighbour interpolation):

$$\hat{I}_t^{(k)} = \varphi_2^{(\text{k})} \circ I_t \ \text{(pdCT)},\quad \hat{S}_t^{(k)} = \varphi_2^{(\text{k})} \circ S_t,\quad \text{k} = 1,2,3 \tag{3}$$

The two steps are complementary: $\varphi_1$ removes inter-patient anatomical differences so that $\varphi_2^{(k)}$ captures change rather than identity, while applying $\varphi_2^{(k)}$ to $I_t$ keeps each prediction anchored to the target patient's own anatomy. Because the foundation model is reused without any patient-specific optimization, the pipeline requires no per-patient training. With one $\varphi_1$ and three $\varphi_2^{(k)}$ per prior patient, the framework yields three pdCT predictions per prior–target pairing.

### 2.4 Dataset and evaluation

**Dataset.** We use a retrospective cohort of 88 HN patients, each with a planning CT (TPCT) and three same-geometry QA CTs (QACT) acquired during treatment; the TPCT is the reference and the QACTs are the treatment-day ground truth. Scans are rigidly aligned to the planning grid (0.977×0.977 mm in-plane, 1.5 mm slices, 512×512 matrix), with up to 45 OAR segmentations per scan (brainstem, cord, parotids, larynx, constrictors, and others). The dataset source is masked for anonymization, with an ethics statement to follow after de-anonymization. This institutional cohort is disjoint from the public data used to pretrain the registration model (Sec. 2.2), so there is no train–test leakage.

**Evaluation protocol.** We evaluate 20 patients as targets, which result in 5,220 pdCTs with 1,740 ($DVF_1$) and 5,220 ($DVF_2$); for each target, prior patients are drawn from the remaining cohort and the target's own QACTs are held out as ground truth. The baseline is *no adaptation*, which uses the planning CT directly on the treatment day and is the current default when no replan is acquired. It is evaluated by comparing $I_t$ against each $Q_t^{(k)}$. The proposed method is evaluated by comparing each pdCT $\hat{I}_t^{(k)}$ against the corresponding QACT. All metrics are computed within a body mask. We report complementary aspects of agreement: NCC and SSIM (intensity and structural agreement) and MAE in HU (intensity error); mean Dice (OAR volumetric overlap), mean HD95 and ASSD in mm (OAR boundary distance), averaged over structures common to the prediction and QACT.

**Composite quality score.** To summarize overall performance with a single number, we map each of six metrics to a composite score with [0,1] scale on which higher is better and average them with equal weights. NCC, SSIM, and Dice already lie in [0,1] and are used directly; the three error metrics are inverse-normalized and clipped:

$$\text{Composite} = \tfrac{1}{6}(\text{NCC}+\text{SSIM}+\text{Dice}+\text{clip}(1-\text{MAE}/M)+\text{clip}(1-\text{HD95}/\delta)+\text{clip}(1-\text{ASSD}/\delta)) \quad (4)$$

where $M = 2000$ HU is the CT intensity range used to normalize MAE, $\delta = 50$ mm caps the boundary distances, and $\text{clip}(x) = \min(\max(x,0),1)$. A higher Composite indicates better overall agreement with the treatment-day QACT.

**Implementation.** Registration hyper-parameters are fixed across all cases (Sec. 2.2). Prior-patient sessions are resampled onto the target grid (linear for intensities, nearest-neighbour for labels) prior to $\varphi_1$. Experiments run on NVIDIA A100 GPUs at roughly 22 min per prediction.

**Ablation.** To test whether the predicted change must be applied to the target's own anatomy, we compare the full method (pdCT) against an ablated variant that removes the longitudinal-change step $\varphi_2^{(k)}$ (pdCT no $\varphi_2$): the $\varphi_1$-aligned prior QACT $\tilde{Q}_p^{(k)}$ is used directly as the prediction, substituting the prior patient's anatomy for the target's (Sec. 3).

# 3 Results and Discussion

## 3.1 Results

**Quantitative agreement.** For each target we evaluate the most challenging treatment-day anatomy, the QACT with the largest change from the planning CT, where no adaptation fails worst and prediction matters most. As Table 1 shows, the predicted CT improves over the no-adaptation baseline on every metric: intensity and structural agreement (NCC, SSIM), mean absolute error (MAE), and OAR overlap and boundary accuracy (Dice, HD95, ASSD), raising the aggregate Composite score from 0.77 to 0.82. Per-patient relative gains exceed 20% for both NCC and Dice.

**Table 1.** Agreement with the treatment-day QACT (largest-change QACT per target), for no adaptation (TPCT) vs. the predicted CT (pdCT). Values are mean ± SD over 20 targets; Improved is the per-patient relative gain.

| | NCC | SSIM | Dice | MAE (HU) | HD95 (mm) | ASSD (mm) | Composite Score |
|---|---|---|---|---|---|---|---|
| TPCT | 0.61±0.12 | 0.83±0.04 | 0.49±0.13 | 110.4±24.3 | 8.9±2.9 | 3.7±1.4 | 0.77±0.06 |
| pdCT | 0.75±0.10 | 0.86±0.03 | 0.59±0.11 | 89.5±16.8 | 7.9±2.8 | 3.1±1.3 | 0.82±0.05 |
| Gain | 22.8% | 2.8% | 20.2% | 23.4% | 10.6% | 17.7% | 6.6% |

*HD = Hausdorff Distance; ASSD = Average Symmetric Surface Distance*

**Image comparisons.** Fig. 2 shows, for a representative case, the planning CT (TPCT), the predicted CT (pdCT), and the treatment-day QACT together with absolute-error maps against the QACT: pdCT visibly reduces residual error in regions of anatomical change while preserving image realism. Fig. 3 overlays propagated OAR contours on the QACT for the TPCT and pdCT conditions, showing improved overlap (per-structure Dice annotated).

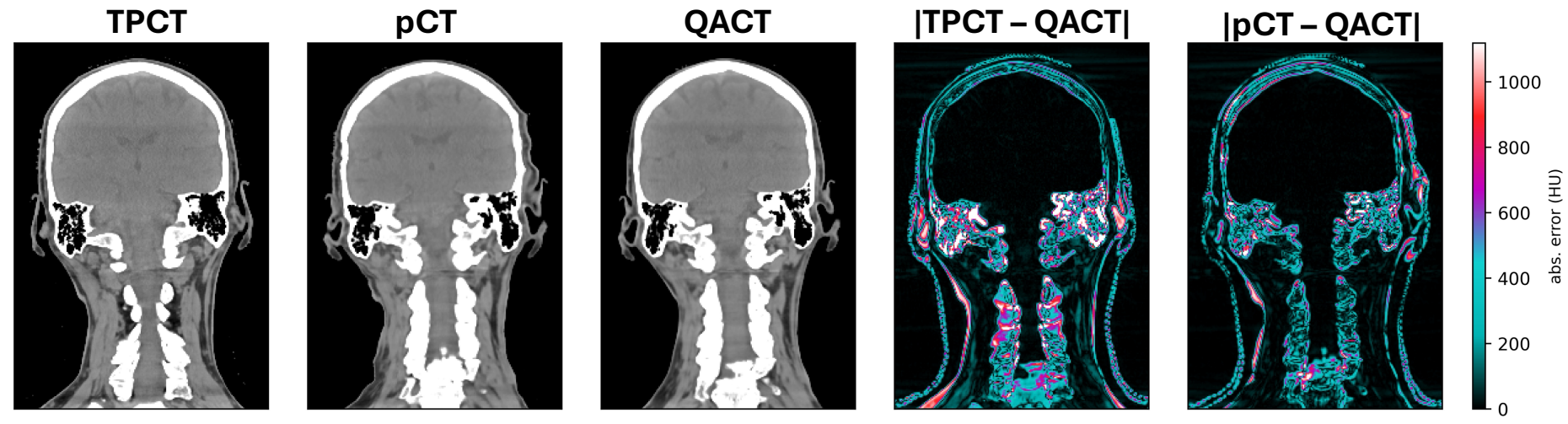

**Fig. 2.** CT quality and error for a representative target patient, on a coronal slice. Columns: planning CT (TPCT), predicted CT (pdCT), treatment-day QACT, and absolute-

error maps |TPCT − QACT | and |pdCT − QACT | on a shared scale. pdCT reduces residual error in regions of anatomical change while preserving image realism.

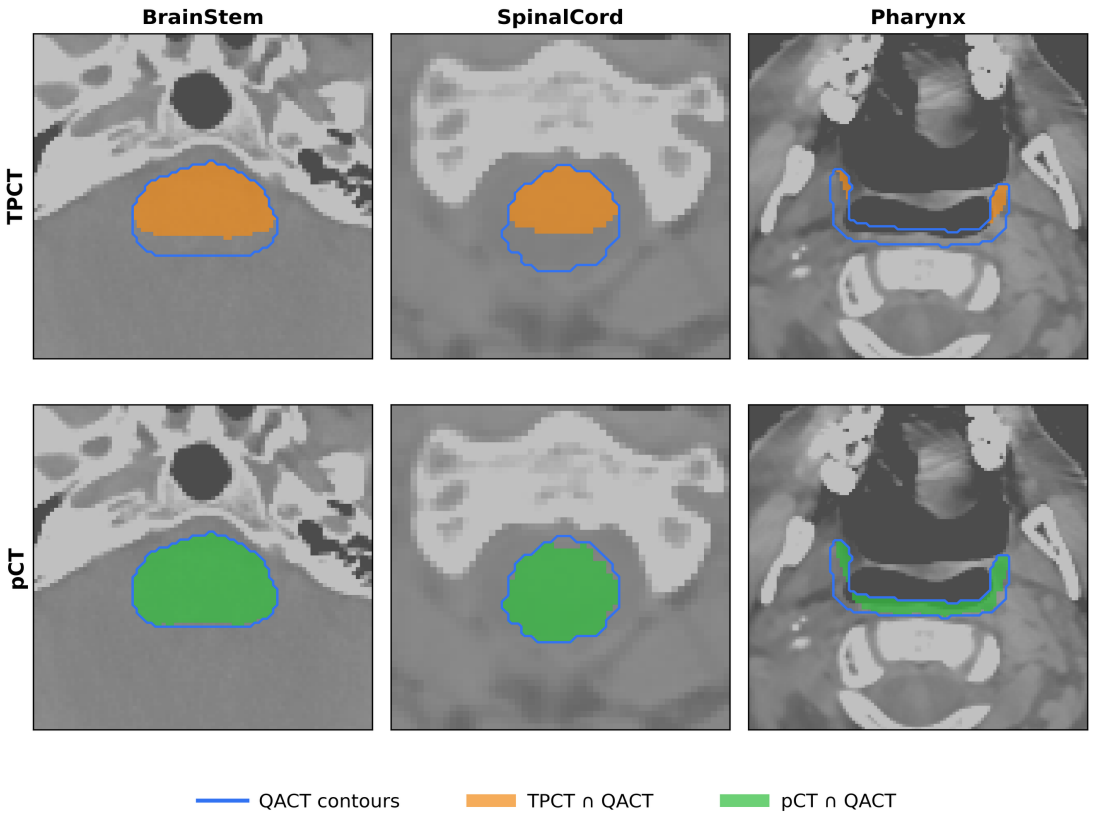

**Fig. 3.** Propagated OAR contours on the treatment-day QACT for a representative target. Rows: TPCT (no adaptation) vs pdCT; columns: brainstem, spinal cord, pharyngeal constrictor. Orange/green = overlap with the QACT for TPCT / pdCT; blue outline = QACT; per-panel Dice annotated.

## 3.2 Ablations

**Removing the longitudinal-change step.** We compare the full method (pdCT) against an ablated variant, pdCT no $\varphi_2$, that skips $\varphi_2$ and instead deformably aligns the prior patient's QACT to the target via $\varphi_1$ and uses it directly, transferring the prior patient's anatomy rather than its longitudinal change (Fig. 4). This variant decreases intensity correlation. It also degrades every structural metric below the full method results since the organ shapes and positions belong to the prior patient's anatomy. Applying the change to the target's own anatomy (full pdCT) is best on every metric, confirming the core design choice: the value comes from transferring the change, not the anatomy.

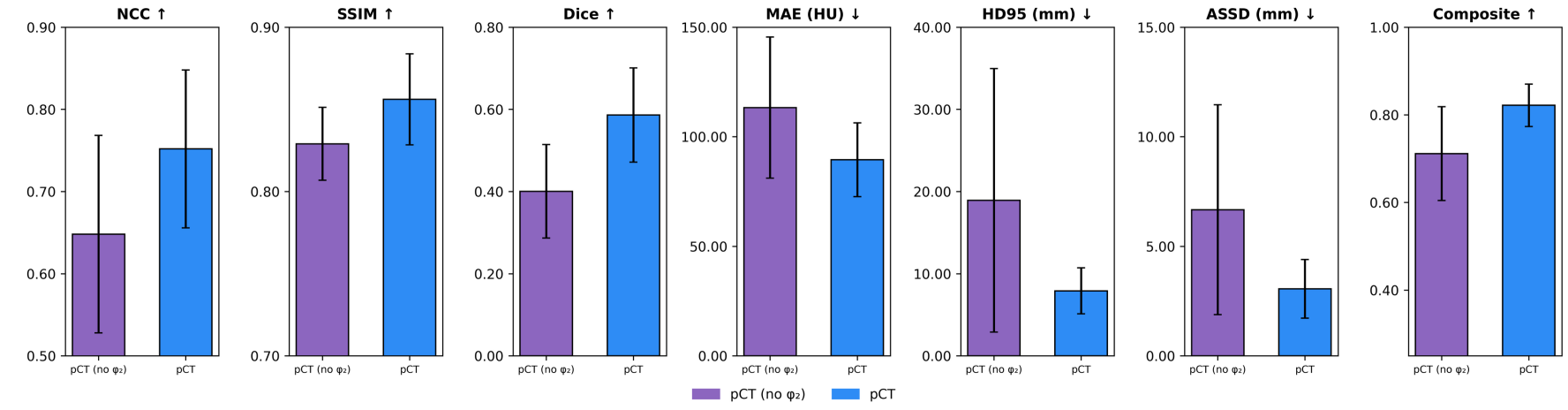

**Fig. 4.** Ablation: full method (pdCT) vs. the variant without the longitudinal-change step (pdCT no $\varphi_2$), mean ± SD over 20 patients. Using the $\varphi_1$-aligned prior QACT directly raises NCC but degrades all structural metrics, while pdCT is best on every metric.

### 3.3 Discussion

The central finding is that a prior patient's planning-to-QACT change, once re-expressed on the target through a single cross-patient registration, predicts the target's own treatment-day anatomy more faithfully than the static planning CT does. This succeeds because anatomical change in HN radiotherapy is largely canonical across patients: tumor and nodal regression, weight-loss-driven neck thinning, and medial parotid migration recur in similar directions and locations, so a longitudinal deformation learned on one patient is informative for another. The two-step design is what makes this usable. The cross-patient field $\varphi_1$ first removes inter-patient anatomical differences so that $\varphi_2^{(k)}$ encodes change rather than identity, and applying $\varphi_2^{(k)}$ to the target's own planning CT keeps the prediction anchored to the target's anatomy. The low folding fraction of $\varphi_2^{(k)}$ confirms that the predicted change is a physically plausible warp rather than a registration artifact.
The ablation isolates the source of the gain. Using the $\varphi_1$-aligned prior QACT directly (pdCT no $\varphi_2$) raises intensity correlation, because a globally aligned prior patient resembles the target in bulk, yet it degrades every structural metric below the no-adaptation baseline, because the borrowed organ shapes and positions belong to the prior patient. Only by transferring the change to the target's own anatomy does the method improve both intensity and structural agreement. The value comes from transferring the change, not the anatomy. This also explains why gains are largest precisely for the patients whose anatomy changed most and negligible when anatomy is stable, and it motivates our evaluation on the largest-change QACT, the case that most needs adaptation.
The dominant remaining bottleneck is prior-patient selection. Our primary results rank candidate priors against the held-out QACT, an oracle unavailable at instantiation time. The composite image-similarity score (NCC, LPIPS, MI) is a step toward a deployable selector, but closing the gap to the oracle ranking will likely require a learned, ground-truth-free scorer that predicts which priors carry transferable change for a given target.
Clinically, the framework reframes adaptation from reactive to proactive: because pdCTs and their contours are synthesized before any treatment-day image is acquired, a precomputed plan library can be selected on the day rather than triggering a multi-day replanning path after drift has occurred. Two limitations bound the study: the evaluation is in the image domain, so dose-domain validation of proton plan quality on the predicted anatomy is needed before clinical use; and the cohort is single-institution and retrospective, so prospective, multi-site validation remains essential.

## 4 Conclusions

This work positions cross-patient longitudinal motion transfer as a viable building block for digital twins in adaptive radiotherapy. Using a single pretrained foundation registration model without any patient-specific training, we synthesize predicted CTs and propagated contours that better match the target's treatment-day anatomy than the static planning CT on every evaluated metric, with physically plausible deformations. The gains are largest for the patients whose anatomy changed most, and an ablation confirms that the benefit comes from transferring the prior patient's longitudinal change to the target's own anatomy rather than borrowing the prior patient's anatomy. Because the prediction is produced before any treatment-day image is acquired, the framework offers a route to shift HN online adaptive proton therapy from reactive to proactive. Learning a ground-truth-free prior-patient selector, together with dose-domain evaluation and prospective validation, is the principal direction for future work.

**Acknowledgments.** This research is supported in part by the National Institutes of Health under Award Number R01DE033512 and R01CA272991.

**Disclosure of Interests.** The authors have no competing interests to declare that are relevant to the content of this article.